\documentclass[a4paper,conference]{IEEEtran}
\usepackage{graphicx,ifthen}
\usepackage{psfrag,amssymb,amsthm}
\usepackage[cmex10]{amsmath}
\usepackage{cite,color,pst-plot,stfloats,pst-sigsys,algorithmicx,algpseudocode}
\usepackage{pstricks-add}

\newtheorem{cor}{Corollary}

\newtheorem{prop}{Proposition}
\theoremstyle{definition}
\newtheorem{definition}{Definition}

\newcommand{\overbar}[1]{\mkern 1.5mu\overline{\mkern-3mu#1\mkern-0.5mu}\mkern 1.5mu}

\newcommand{\ent}[1]{H(#1)}

\newcommand{\entrate}[1]{\overbar{H}(\mathbf{#1})}

\newcommand{\loss}[2][\empty]{\ifthenelse{\equal{#1}{\empty}}{L(#2)}{L_{#1}(#2)}}
\newcommand{\lossrate}[2][\empty]{\ifthenelse{\equal{#1}{\empty}}{\overbar{L}(\mathbf{#2})}{L_{\mathbf{#1}}(\mathbf{#2})}}
\newcommand{\relLoss}[2][\empty]{\ifthenelse{\equal{#1}{\empty}}{l(#2)}{l_{#1}(#2)}}

\newcommand{\dom}[1]{\mathcal{#1}}

\newcommand{\Prob}[1]{\mathrm{Pr}(#1)}

\newcommand{\card}[1]{\mathrm{card}(#1)}

\newcommand{\xvec}{\mathbf{x}}
\newcommand{\Xvec}{\mathbf{X}}

\newcommand{\Yvec}{\mathbf{Y}}
\newcommand{\yvec}{\mathbf{y}}

\newcommand{\Pvec}{\mathbf{P}}
\newcommand{\muvec}{\boldsymbol{\mu}}

\newcommand{\Amat}{\mathbf{A}}

\newcommand{\preim}[1]{g^{-1}[#1]}

\newcommand{\limn}{\lim_{n\to\infty}}

\newcommand{\charl}[1]{`\texttt{#1}'}
\newcommand{\preimCnt}{T_n}

\newcommand{\zvec}{\mathbf{z}}

\def \arxiv {1}

\title{Information-Preserving Markov Aggregation}

\author{
\IEEEauthorblockN{Bernhard C. Geiger\IEEEauthorrefmark{1}, Christoph Temmel\IEEEauthorrefmark{2}}
\IEEEauthorblockA{
\IEEEauthorrefmark{1}Signal Processing and Speech Communication Laboratory, Graz University of Technology, Austria\\
\IEEEauthorrefmark{2}Department of Mathematics, VU University - Faculty of Sciences, Amsterdam, The Netherlands\\
geiger@ieee.org, ctc@temmel.me}
}

\begin{document}

\maketitle

\begin{abstract}
We present a sufficient condition for a non-injective function of a Markov chain to be a second-order Markov chain with the same entropy rate as the original chain. This permits an information-preserving state space reduction by merging states or, equivalently, lossless compression of a Markov source on a sample-by-sample basis. The cardinality of the reduced state space is bounded from below by the node degrees of the transition graph associated with the original Markov chain.

We also present an algorithm listing all possible information-preserving state space reductions, for a given transition graph. We illustrate our results by applying the algorithm to a bi-gram letter model of an English text.
\end{abstract}

\begin{IEEEkeywords}
lossless compression, Markov chain, model order reduction, $n$-gram model
\end{IEEEkeywords}

\section{Introduction}\label{sec:Intro}
Markov chains are ubiquitously used in many scientific fields, ranging from machine learning and systems biology over speech processing to information theory, where they act as models for sources and channels. In some of these fields, however, the state space of the Markov chain is too large to allow either proper training of the model (see $n$-grams in speech processing~\cite{Manning_NLP}) or its simulation (as in chemical reaction networks~\cite{Wilkinson_SystemsBiology}). In these situations it is convenient to define a model of the process on a smaller state space, which not only allows efficient simulation (such as higher-order Markov models), but also preserves as much information of the original model as possible.

One way to reduce the cardinality of the state space of a Markov chain is to merge states, which is equivalent to feeding the process through a non-injective function. The merging usually depends on the cost function; candidate methods either rely on the Fiedler vector or other spectral criteria~\cite{Meila_Segmentation,Runolfsson_ModelReduction}, or on the Kullback-Leibler divergence rate w.r.t. some reference process~\cite{Meyn_MarkovAggregation,Vidyasagar_MarkovAgg}.

In addition to the model information lost by merging, the obtained process does, in general, not possess the Markov property. Modeling it as a Markov chain on the reduced state space as suggested in, e.g.,~\cite{Meyn_MarkovAggregation}, typically leads to an additional loss of model information. The same holds for Markov models obtained from clustered training data, as, e.g., for the $n$-gram class model in~\cite{Brown_NGrams}. Consequently, there is a trade-off between cardinality of the state space, model complexity, and information loss.

Recently, we have shown the existence of sufficient conditions on a Markov chain and a non-injective function merging its states such that the obtained process is not only a $k$th-order Markov chain (which is desirable from a computational point-of-view), but also preserves full model information~\cite{GeigerTemmel_kLump}. While the former property is commonly referred to as \emph{lumpability}, the latter is a rather surprising one: whereas, in principle, stationary sources can be compressed efficiently by assigning codewords to \emph{blocks} of samples, our result shows that in some cases lossless compression is possible on a sample-by-sample basis: Encoding is trivial, and the decoder only needs to remember the last $k$ symbols.

Extending our previous results, we show in Section~\ref{sec:prevExt}, using spectral theory of graphs, that for an information-preserving compression, the number of input sequences merged to the same output sequence is bounded independently of the sequence length. This result allows us to estimate the minimum cardinality of the reduced state space based on the degree structure of the transition graph of the original Markov chain. Furthermore, we prove that, if a specific partition of the original state space satisfies the sufficient conditions for $k$th-order Markovity and information-preservation, then so does every refinement of this partition. Section~\ref{sec:sfs2} focuses on second-order Markov chains, due to their computationally desirable properties, and presents an iterative algorithm listing all possible partitions satisfying the abovementioned sufficient conditions. 
\ifthenelse{\arxiv=1}
{To illustrate the algorithm, we introduce a simple toy example in Section~\ref{sec:toyExample} before analyzing a bi-gram letter model in Section~\ref{sec:ngram}.}
{We finally apply our algorithm to a bi-gram letter model in Section~\ref{sec:ngram}.

An extended version of this paper can be found in~\cite{Geiger_Markov_arXiv}.}

\section{Preliminaries \& Notation}\label{sec:prelim}
Throughout this work, we deal with an irreducible, aperiodic, homogeneous Markov chain $\Xvec$ on a finite state space $\dom{X}$ and with transition matrix $\Pvec$. Let $X_n$ be the $n$th sample of the process, and let $X_i^j:=\{X_i,X_{i+1},\dots,X_j\}$. We assume that $\Xvec$ is stationary, i.e., that the initial distribution of the chain coincides with its invariant distribution $\muvec$. Hence, for every $n$, the distribution $P_{X_n}$ of $X_n$ equals $\muvec$.

We consider a surjective \emph{lumping function} $g{:}\ \dom{X}\to\dom{Y}$, with $\card{\dom{X}}=:N>M:=\card{\dom{Y}}\ge 2$. Abusing notation, we extend $g$ to $\dom{X}^n\to\dom{Y}^n$ coordinate-wise and denote by $\preim{y}$ the preimage of $y$ under $g$. We call the stationary stochastic process $\Yvec$, defined by $Y_n:=g(X_n)$, the \emph{lumped process} and the tuple $(\Pvec,g)$ the \emph{lumping}.

Since the lumping function is non-injective, a loss of information may occur, which we quantify by the conditional entropy rate
\begin{equation}
	\entrate{X|Y}:=\limn \frac{1}{n}\ent{X_1^n|Y_1^n} = \entrate{X}-\entrate{Y}
\end{equation}
where $\ent{\cdot}$ and $\entrate{\cdot}$ denote the entropy and the entropy rate (if it exists) of the argument, respectively. The lumping $(\Pvec,g)$ is \emph{information-preserving} iff $\entrate{X|Y}=0$.

\section{Previous Results \& Extensions}\label{sec:prevExt}
We summarize several definitions and results from~\cite{GeigerTemmel_kLump} relevant to this work:
\begin{definition}[Preimage Count]\label{def:preimageCnt}
The \emph{preimage count of length $n$} is the random variable
\begin{equation}
	\preimCnt:= \sum_{\xvec\in\preim{Y_1^n}} \left[\Prob{X_1^n=\xvec}>0 \right]
\end{equation}
where $\left[ A\right]=1$ if $A$ is true and zero otherwise (Iverson bracket).
\end{definition}

In other words, the preimage count maps each sequence of length $n$ of the output process $\Yvec$ to the cardinality of the realizable portion of its preimage.

The following characterization holds~\cite[Thm.~1]{GeigerTemmel_kLump}:
\begin{subequations}
	\begin{align}
		\entrate{X|Y}=0 &\Leftrightarrow \exists\, C<\infty{:}\ \Prob{\sup_{n\to\infty} \preimCnt \le C}=1 \label{eq:OrigTrajPreservation}\\
		\entrate{X|Y}>0 &\Leftrightarrow \exists\, C>1{:}\ \Prob{\liminf_{n\to\infty} \sqrt[n]{\preimCnt} \ge C}=1 \label{eq:OrigTrajExplosion}
	\end{align}
\end{subequations}
i.e., that an almost-surely bounded preimage count (for arbitrary sequence length $n$) is equivalent to a vanishing information loss rate. 

The information-preserving case~\eqref{eq:OrigTrajPreservation} can be strengthened to a deterministic version:
\begin{prop}[Bounded Preimage Count]\label{prop:TrajPres}
\begin{equation}
	\entrate{X|Y}=0\Leftrightarrow \exists\, C<\infty{:}\  \sup_{n\to\infty} \preimCnt \le C\,.
\end{equation}
\end{prop}

\begin{IEEEproof}
 See Appendix.
\end{IEEEproof}

An interesting line for future research would be to show a deterministic analog of~\eqref{eq:OrigTrajExplosion} and its direct derivation from the Shannon-McMillan-Breiman theorem~\cite[Ch.~16.8]{Cover_Information2}.

As a corollary to Proposition~\ref{prop:TrajPres} we get
\begin{cor}\label{cor:Mbound}
An information-preserving lumping $(\Pvec,g)$ satisfies
\begin{equation}
	M \ge \min_i d_i
\end{equation}
where $d_i:=\sum_{j=1}^N \left[ P_{i,j}>0\right]$ is the out-degree of state $i$.
\end{cor}

\begin{IEEEproof}
See Appendix.
\end{IEEEproof}

Corollary~\ref{cor:Mbound} upper-bounds the possible state space reduction of an information-preserving lumping. In particular, a Markov chain with a positive transition matrix $\Pvec$ does not admit an information-preserving lumping~\cite[Cor.~4]{GeigerTemmel_kLump}: In this case, all states have out-degree $N$, and the bound $M\ge N$ only holds for the trivial lumping.

Complementing this necessary condition for preservation of information, in~\cite[Prop.~10]{GeigerTemmel_kLump} we also gave a sufficient condition, additionally implying that $\Yvec$ is a $k$th-order Markov chain, i.e., that $\forall n: \forall x_1^n\in\dom{X}^n:$
\begin{equation}
	\Prob{X_n=x_n|X_1^{n-1}=x_1^{n-1}} = \Prob{X_n=x_n|X_{n-k}^{n-1}=x_{n-k}^{n-1}}\,.
\end{equation}
To this end, we introduced

\begin{definition}[Single Forward Sequence{~\cite[Def.~9]{GeigerTemmel_kLump}}]\label{def:SFSk}
For $k\ge 2$ a lumping $(\Pvec,g)$ has the \emph{single forward $k$-sequence} property (short: $\mathsf{SFS}(k)$) iff
\begin{multline}
	\forall \yvec\in\dom{Y}^{k-1},y\in\dom{Y}{:}\ \exists!\xvec'\in\preim{\yvec}{:}\\
	\forall x\in\preim{y},\xvec\in\preim{\yvec}\setminus\{\xvec'\}{:}\\
	\Prob{X_2^{k}=\xvec|Y_2^{k}=\yvec,X_1=x}=0\,.
\end{multline}
\end{definition}

Thus, for every realization of $Y_1^n$, the realizable preimage of $Y_2^n$ is a singleton. Therefore, $\mathsf{SFS}(k)$ implies not only that $\Yvec$ is $k$th-order Markov, but also that the lumping is information-preserving\footnote{Actually, $\mathsf{SFS}(k)$ implies more than $\entrate{X|Y}=0$: It implies that a sequence of states of the reduced model uniquely determines the corresponding sequence of the original model, except for the first sample. Thus, the reduced model is in some sense ``invertible''.}~\cite[Prop.~10]{GeigerTemmel_kLump}. It is a property of the combinatorial structure of the transition matrix $\Pvec$, i.e., it only depends on the location of its non-zero entries, and can be checked with a complexity of $\mathcal{O}(N^k)$~\cite{GeigerTemmel_kLump}.

The $\mathsf{SFS}(k)$-property has practical significance: Besides preserving, if possible, the information of the original model, those lumpings which possess the Markov property of any (low) order are preferable from a computational perspective. Moreover, the corresponding conditions for the more desirable first-order Markov output, not necessarily information-preserving, are too restrictive in most scenarios (cf.~\cite[Sec.~6.3]{Kemeny_FMC}).

The next result investigates a cascade of lumpings. Below, we identify a function with the partition it induces on its domain. Let $h{:}\ \dom{X}\to\dom{Z}$, $f{:}\ \dom{Z}\to\dom{Y}$, and $g:=h\circ f$ be $\dom{X}\to\dom{Y}$. Clearly, (the partition induced by) $g$ is coarser than (the partition induced by) $h$ because of the intermediate application of $f$. In other words, $h$ is a \emph{refinement} of $g$. 
\begin{prop}[$\mathsf{SFS}(k)$ \& Refinements]\label{prop:SFSRefined}
If a lumping $(\Pvec,g)$ is $\mathsf{SFS}(k)$, then so is $(\Pvec,h)$, for all refinements $h$ of $g$.
\end{prop}

\begin{IEEEproof}
See Appendix.
\end{IEEEproof}

A refinement does not increase the loss of information, so information-preservation is preserved under refinements. In contrast, a refinement of a lumping yielding a $k$th-order Markov process $\Yvec$ need not possess that property; the lumping to a single state has the Markov property, while a refinement of it generally has not. 
\ifthenelse{\arxiv=1}
{All $\mathsf{SFS}(k)$-lumpings lie within the intersection of information-preserving lumpings and lumpings yielding a $k$th-order Markov chain. However, as shown in~\cite{GeigerTemmel_kLump}, the $\mathsf{SFS}(k)$-property does not exhaust this intersection.}
{}

\section{An Algorithm for $\mathsf{SFS}(2)$-lumpings}\label{sec:sfs2}
In this section, we present an algorithm listing all $\mathsf{SFS}(2)$-lumpings, i.e., lumpings $(\Pvec,g)$ yielding a second-order Markov chain and preserving full model information. There are two reasons for focusing on this particular class of lumpings: Firstly, low-order Markov models are attractive from a computational point-of-view. To obtain a first-order Markov model, the transition chain of the original chain has to satisfy overly restrictive conditions. Thus, a second-order model represents a good trade-off between computational efficiency and applicability. Secondly, compared to the general case, whether a lumping $(\Pvec,g)$ satisfies the $\mathsf{SFS}(2)$-property can be determined by looking only at the transition matrix\footnote{This does not conflict with the statement, that \emph{in general} the $\mathsf{SFS}(k)$-property depends on the combinatorial structure of $\Pvec$: For $k>2$ this dependency is more complicated than for $k=2$.} $\Pvec$. $\mathsf{SFS}(2)$-lumpings have the property that, for all $y_1,y_2\in\dom{Y}$, from within a set $\preim{y_1}$ at most one element in the set $\preim{y_2}$ is accessible:
\begin{multline}
	\forall y_1,y_2\in\dom{Y}{:}\ \exists!\,x_2'\in\preim{y_2}{:}\\
	\forall x_1\in\preim{y_1},x_2\in\preim{y_2}\setminus\{x_2'\}{:}\quad 
	P_{x_1,x_2}=0\,.\label{eq:sfs2}
\end{multline}
This gives rise to
\begin{prop}\label{prop:Msfs2}
An $\mathsf{SFS}(2)$-lumping satisfies
\begin{equation}
	M\ge \max_i d_i\,.
\end{equation}
\end{prop}

\begin{IEEEproof}
We evaluate the rows of $\Pvec$ separately. All states $x_2$ accessible from state $x_1$ are characterized by $P_{x_1,x_2}>0$. Any two states accessible from $x_1$ cannot be merged, since this would contradict~\eqref{eq:sfs2}. Thus, all states accessible from $x_1$ must have different images, implying $M\ge d_{x_1}$. The result follows by considering all states $x_1$.
\end{IEEEproof}

In particular, Proposition~\ref{prop:Msfs2} implies that a transition matrix with at least one positive row does not admit an $\mathsf{SFS}(2)$-lumping.

An algorithm listing all $\mathsf{SFS}(2)$-lumpings, or $\mathsf{SFS}(2)$-partitions, for a given transition matrix $\Pvec$ has to check the $\mathsf{SFS}(2)$-property for all partitions of $\dom{X}$ into at least $\max_i d_i$ non-empty sets. The number of these partitions can be calculated from the Stirling numbers of the second kind~\cite[Thm.~8.2.5]{Brualdi_Combinatorics} and is typically too large to allow an exhaustive search. Therefore, we use Proposition~\ref{prop:SFSRefined} to reduce the search space.

Starting from the trivial partition with $N$ blocks, we evaluate all possible merges of two states, i.e., all possible partitions with $N-1$ sets, of which there exist $\frac{N(N-1)}{2}$. Out of these, we drop those from the list which do not possess the $\mathsf{SFS}(2)$-property. The remaining set of \emph{admissible pairs} is a central element of the algorithm.

We proceed iteratively: To generate all candidate partitions with $N-i$ sets, we perform all admissible pair-wise merges on all $\mathsf{SFS}(2)$-partitions with $N-i+1$ sets. An admissible pair-wise merge is a merge of two sets of a partition, where either set contains one element of the admissible pair. From the resulting partitions one drops those violating $\mathsf{SFS}(2)$ before performing the next iteration. Since this algorithm generates some partitions multiple times (\ifthenelse{\arxiv=1}{see the toy example in Section~\ref{sec:toyExample}}{see the toy example in~\cite{Geiger_Markov_arXiv}}), in every iteration all duplicates are removed. The algorithm is presented in Table~\ref{tab:algo}.

Iterative generation of the partitions by admissible pair-wise merges allows application of Proposition~\ref{prop:SFSRefined}, which reduces the number of partitions to be searched. If the number of admissible pairs is small compared to $\frac{N(N-1)}{2}$, then this reduction is significant. Inefficiencies in our algorithm caused by multiple considerations of the same partitions could be alleviated by adapting the classical algorithms for the partition generating problem~\cite{Er_Partitions,Semba_Partitions}.

The actual choice of one of the obtained $\mathsf{SFS}(2)$-partitions for model order reduction requires additional model-specific considerations: A possible criterion could be maximum compression (i.e., smallest entropy of the marginal distribution). The toy example in~\cite{Geiger_Markov_arXiv}, for instance, illustrates the case where compression using $\mathsf{SFS}(2)$-lumpings is optimal, i.e., where $\ent{Y}=\entrate{\Yvec}=\entrate{\Xvec}$.

In practice, if $N$ is large and if the number of admissible pairs is of the same order as $\frac{N(N-1)}{2}$, listing all $\mathsf{SFS}(2)$-partitions might be computationally expensive. One can trade optimality for speed by greedy selection of the best $h$ in line~\ref{algstep:lumps}, given a specific criterion, or by evaluating only a (random) subset of admissible pairs in line~\ref{algstep:innermost}.

\begin{table}
\caption{Algorithm for listing all $\mathsf{SFS}(2)$-lumpings}\label{tab:algo}
\hrule
 \begin{algorithmic}[1]
\Function{ListLumpings}{$\Pvec$}
\State $\text{admPairs}\gets\textsc{GetAdmissiblePairs}(\Pvec)$
\State $\text{Lumpings}(1)\gets\text{merge}(\text{admPairs})$ \Comment{Convert pairs to functions}
\State $n\gets 1$
\While{$\text{notEmpty}(\text{Lumpings}(n))$}
	\State $n\gets n+1$
	\State $\text{Lumpings}(n)\gets [\ ]$
	\For {$h\in\text{Lumpings}(n-1)$}\label{algstep:lumps}
\For{$\{i_1,i_2\}\in\text{admPairs}$}\label{algstep:innermost}
	\State $g\gets h$
	\State $g(h^{-1}(h(i_2)))\gets g(i_1)$ \Comment{$i_1$ and $i_2$ have same image.}
	\If{$g \text{ is } \mathsf{SFS}(2)$}
		\State $\text{Lumpings}(n) \gets [\text{Lumpings}(n); g]$
	\EndIf
\EndFor
	\EndFor
	\State Remove duplicates from $\text{Lumpings}$
\EndWhile
\State \Return $\text{Lumpings}$
\EndFunction
\Statex
\hrule
\Function{GetAdmissiblePairs}{$\Pvec$}
\State $\text{Pairs}\gets [\ ]$
\State $N\gets \text{dim}(\Pvec)$
\For{$i_1=1:N-1$}
	\For{$i_2=i_1:N$}
		\State $f\gets \text{merge}(i_1,i_2)$ \Comment{$f$ merges $i_1$ and $i_2$}
		\If{$f \text{ is } \mathsf{SFS}(2)$}
			\State $\text{Pairs} \gets [\text{Pairs}; \{i_1,i_2\}]$
		\EndIf
	\EndFor
\EndFor
\State \Return $\text{Pairs}$
\EndFunction
\end{algorithmic}
\hrule
\end{table}

\ifthenelse{\arxiv=1}{
\section{A Toy Example}\label{sec:toyExample}

\begin{figure}
	\centering
	\begin{pspicture}[showgrid=false](0.2,-0.7)(0.5,5)
	\psmatrix[mnode=circle,emnode=R,colsep=1.5,rowsep=1.5,fillstyle=solid]
	&    [name=6]6 \\
	&  & [name=1,fillcolor=red!30]  $1$  & [name=2,fillcolor=red!30] $2$ & [name=3,fillcolor=red!30] $3$ \\
	&    [name=4,fillcolor=blue!30] $4$  & [name=5,fillcolor=blue!30] $5$
	\endpsmatrix
	\psset{shortput=nab,arrows=->,labelsep=3pt,nrot=:U}
	\small
	\ncline{1}{2}
	\nccircle[angle=180]{2}{0.3}
	\ncline{3}{2}
	\nccircle{6}{0.3}
	\ncline{6}{1}
	\ncarc[arcangle=-40]{6}{4}
	\ncarc[arcangle=0]{4}{6}
	\ncline{5}{4}
	\nccircle[angle=120]{4}{0.3}
	\ncarc[arcangle=0,arrows=<-]{6}{5}
	\ncarc[arcangle=0]{1}{5}
	\ncarc[arcangle=20,arrows=<-]{5}{2}
	\ncarc[arcangle=0,arrows=<-]{5}{3}
	\ncarc[arcangle=-20]{5}{3}
	\ncarc[arcangle=30,arrows=<-]{6}{1}
	\ncarc[arcangle=-50]{4}{3}
	\ncarc[arcangle=30,arrows=<-]{6}{2}
	\ncarc[arcangle=30,arrows=<-]{6}{3}
	\psellipse[linestyle=dashed,linecolor=red](-2.3,2.0)(2.7,0.9)
	\psellipse[linestyle=dashed,linecolor=blue](-5.5,-.0)(1.9,0.9)
\end{pspicture}%
	\caption{A transition graph on six vertices with a lumping given by the partition $\{\textcolor{red}{\{1,2,3\}},\textcolor{blue}{\{4,5\}},\{6\}\}$. The lumping is of type $\mathsf{SFS}(2)$.}
	\label{fig:toyExample}
\end{figure}
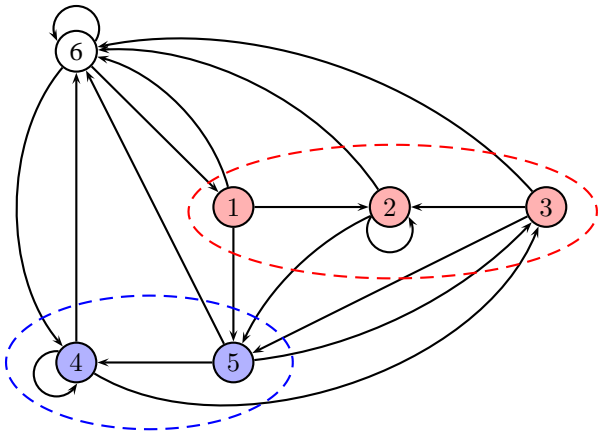

\begin{figure*}
	\centering
	\begin{pspicture}[showgrid=false](0.5,-0.5)(18,5)
	\small
	\psmatrix[mnode=R,emnode=R,colsep=0.75,rowsep=2,fillstyle=solid]
		& [name=12]$\{1,2\}$ && [name=13]$\{1,3\}$ && [name=15]$\{1,5\}$ && [name=23]$\{2,3\}$ & [name=45]$\{4,5\}$\\%
		&[name=123]$\{1,2,3\}$ & [name=125]\textcolor{gray}{$\{1,2,5\}$} & [name=12and45]$\{1,2\},\{4,5\}$ & [name=135]\textcolor{gray}{$\{1,3,5\}$} & [name=13and45]$\{1,3\},\{4,5\}$  & [name=15-23]\textcolor{gray}{$\{1,5\},\{2,3\}$} & [name=145]\textcolor{gray}{$\{1,4,5\}$}& [name=23and45]$\{2,3\},\{4,5\}$\\%
		&[name=1235]\textcolor{gray}{$\{1,2,3,5\}$} && [name=123and45]$\{1,2,3\},\{4,5\}$ &[name=1245]\textcolor{gray}{$\{1,2,4,5\}$} &[name=1345]\textcolor{gray}{$\{1,3,4,5\}$}&&&[name=23and145]\textcolor{gray}{$\{2,3\},\{1,4,5\}$}\\%
	\endpsmatrix
	\psset{shortput=nab,arrows=->,labelsep=3pt,nrot=:U}
	\ncline[linecolor=red,linewidth=1.5pt]{12}{123}
	\ncline[linecolor=red,linewidth=1.5pt]{12}{125}
	\ncline[linecolor=red,linewidth=1.5pt]{12}{12and45}
	\ncline[linecolor=red,linewidth=1.5pt]{13}{135}
	\ncline[linecolor=red,linewidth=1.5pt]{13}{13and45}
	\ncline[linecolor=red,linewidth=1.5pt]{15}{145}
	\ncline[linecolor=red,linewidth=1.5pt]{15}{15and23}
	\ncline[linecolor=red,linewidth=1.5pt]{23}{23and45}
	\ncline[linecolor=gray,linewidth=0.5pt]{13}{123}
	\ncline[linecolor=gray,linewidth=0.5pt]{15}{125}
	\ncline[linecolor=gray,linewidth=0.5pt]{15}{135}
	\ncline[linecolor=gray,linewidth=0.5pt]{23}{123}
	\ncline[linecolor=gray,linewidth=0.5pt]{23}{15and23}
	\ncline[linecolor=gray,linewidth=0.5pt]{45}{12and45}
	\ncline[linecolor=gray,linewidth=0.5pt]{45}{13and45}
	\ncline[linecolor=gray,linewidth=0.5pt]{45}{23and45}
	\ncline[linecolor=gray,linewidth=0.5pt]{45}{145}
	\ncline[linecolor=red,linewidth=1.5pt]{123}{1235}
	\ncline[linecolor=red,linewidth=1.5pt]{123}{123and45}
	\ncline[linecolor=red,linewidth=1.5pt]{12and45}{1245}
	\ncline[linecolor=red,linewidth=1.5pt]{13and45}{1345}
	\ncline[linecolor=red,linewidth=1.5pt]{23and45}{23and145}
	\ncline[linecolor=gray,linewidth=0.5pt]{12and45}{123and45}
	\ncline[linecolor=gray,linewidth=0.5pt]{23and45}{123and45}
	\ncline[linecolor=gray,linewidth=0.5pt]{13and45}{123and45}
\end{pspicture}%
	\caption{An illustration of the algorithm of Table~\ref{tab:algo} at the hand of the example depicted in Fig.~\ref{fig:toyExample}. The first row shows all admissible pairs, the algorithm runs through all rows (top to bottom) by merging according to the admissible pairs (left to right). Bold, red arrows indicate newly generated partitions, gray arrows indicate that this partition was already found and is thus removed as a duplicate. Gray partitions violate the $\mathsf{SFS}(2)$-property. This figure lists all $\mathsf{SFS}(2)$-partitions of $\dom{X}$ (cf.~Table~\ref{tab:toyExample}).}
	\label{fig:algoIllustration}
\end{figure*}
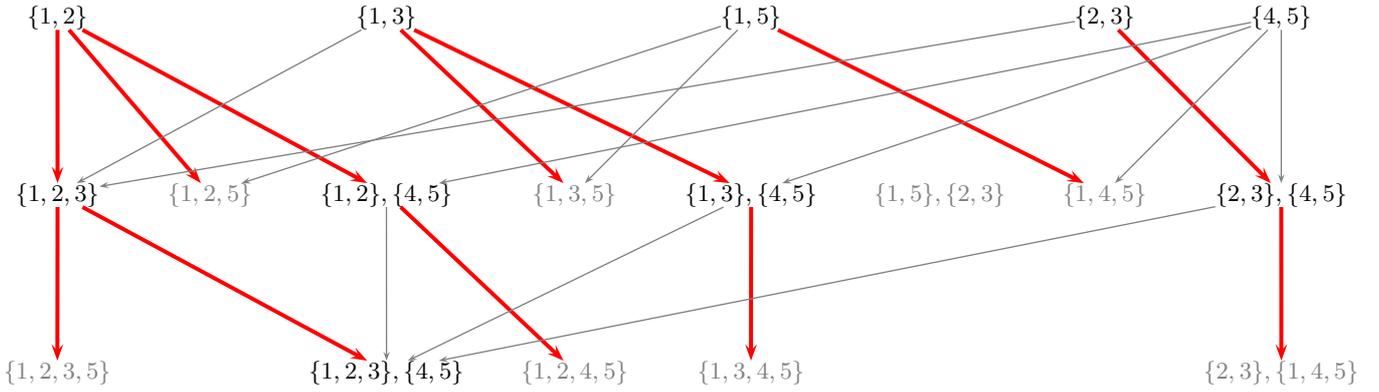

We illustrate our algorithm at the hand of a small example. Consider the six-state Markov chain with transition graph depicted in Fig~\ref{fig:toyExample}, whose adjacency matrix $\Amat$ is
\begin{equation}
	\Amat = \left[
		\begin{array}{cccccc}
		0&	1&	0&	0&	1&	1\\
		0&	1&	0&	0&	1&	1\\
		0&	1&	0&	0&	1&	1\\
		0&	0&	1&	1&	0&	1\\
		0&	0&	1&	1&	0&	1\\
		1&	0&	0&	1&	0&	1
		\end{array}
	\right]\,.
\end{equation}
Since all states have out-degree $d_i=3$, lumpings to at least $M=3$ states are considered. The lumping in Fig.~\ref{fig:toyExample} satisfies the $\mathsf{SFS}(2)$-property. Fig.~\ref{fig:algoIllustration} shows the derivation of the lumping of Fig.~\ref{fig:toyExample} by our algorithm.

During initialization we evaluate all 15 possible pair-wise merges. Of these, we exclude all pairs where both members are accessible from the same state, i.e., $\{2,5\}$, $\{2,6\}$, $\{5,6\}$, $\{3,4\}$, $\{3,6\}$, $\{4,6\}$, $\{1,4\}$, and $\{1,6\}$. Furthermore, $\{2,4\}$ and $\{3,5\}$ are excluded too; the former because both states have self-loops, the latter because both states are connected in either direction. Only five pairs are admissible.

One admissible pair is $\{1,2\}$, i.e., the function $h$ merging $\{1,2\}$ and, thus, inducing the partition $\dom{Z}_5=\{\{1,2\},\{3\},\{4\},\{5\},\{6\}\}$, satisfies $\mathsf{SFS}(2)$. With this $h$ we enter the algorithm in the innermost loop (Table~\ref{tab:algo}, line~\ref{algstep:innermost}). The algorithm performs pair-wise merges according to the five admissible pairs and obtains the following merges: $\{1,2\}$, $\{1,2,3\}$, $\{1,2,5\}$, $\{\{1,2\},\{4,5\}\}$; the first is a (trivial) duplicate (by performing a pair-wise merge according to $\{1,2\}$) and the second is obtained twice (by pairing $\{1,2\}$ with $\{1,3\}$ and $\{2,3\}$). Only $\{1,2,5\}$ violates $\mathsf{SFS}(2)$. The functions merging $\{1,2,3\}$ and $\{\{1,2\},\{4,5\}\}$ are added to the list of lumping functions to four states, and the procedure is repeated for a different admissible pair.

For the next iteration, fix $h$ such that it induces the partition $\dom{Z}_4=\{\{1,2,3\},\{4\},\{5\},\{6\}\}$. The five admissible pairs yield the non-trivial merges $\{1,2,3\}$, a duplicate which is obtained three times, $\{1,2,3,5\}$, which violates $\mathsf{SFS}(2)$, and $\{\{1,2,3\},\{4,5\}\}$, which is the solution depicted in Fig.~\ref{fig:toyExample}. The algorithm terminates now, since every pair-wise merge of $\dom{Z}_3=\dom{Y}=\{\{1,2,3\},\{4,5\},\{6\}\}$ either violates $\mathsf{SFS}(2)$ or is a duplicate. The list of all $\mathsf{SFS}(2)$-lumpings found by the algorithm is given in Table~\ref{tab:toyExample}.

Interestingly, if for the given transition graph all transition probabilities are set to $1/3$, it can be shown that the lumped process $\Yvec$ is a sequence of iid random variables. This observation does not conflict with the $\mathsf{SFS}(2)$-property, since an iid process is Markov of every order. Furthermore, since the lumping is information-preserving and since the redundancy of $\Yvec$ vanishes, one has $\ent{Y}=\entrate{\Yvec}$. The compression achieved by this simple symbol-by-symbol encoding is optimal for this example.

\begin{table}
\caption{List of $\mathsf{SFS}(2)$-lumpings of the toy example found by the algorithm}
\label{tab:toyExample}
\centering
	\begin{tabular}{l|c}
	$M$ & Partition $\dom{Z}_M$\\
	\hline
	6 & $\{1\},\{2\},\{3\},\{4\},\{5\},\{6\}$\\
	\hline
	5 & $\{1,2\},\{3\},\{4\},\{5\},\{6\}$\\
	& $\{1,3\},\{2\},\{4\},\{5\},\{6\}$\\
	& $\{1,5\},\{2\},\{3\},\{4\},\{6\}$\\
	& $\{1\},\{2,3\},\{3\},\{4\},\{6\}$\\
	& $\{1\},\{2\},\{3\},\{4,5\},\{6\}$\\
	\hline
	4 & $\{1,2,3\},\{4\},\{5\},\{6\}$\\
	& $\{1,2\},\{3\},\{4,5\},\{6\}$\\
	& $\{1,3\},\{2\},\{4,5\},\{6\}$\\
	& $\{1\},\{2,3\},\{4,5\},\{6\}$\\
	\hline
	3 & $\{1,2,3\},\{4,5\},\{6\}$
	\end{tabular}
\end{table}
}
{}

\section{Clustering a bi-gram model}\label{sec:ngram}

We apply our algorithm to a bi-gram\footnote{Shannon used bi-grams, or \emph{digrams} as he called them, as a second-order approximation of the English language~\cite{Shannon_TheoryOfComm}.} letter model. 
Commonly used in speech processing~\cite[Ch.~6]{Manning_NLP}, $n$-grams (of which bi-grams are a special case) are $(n-1)$th-order Markov models for the occurrence of letters or words.
From a set of training data the relative frequency of the (co-)occurrence of letters or words is determined, yielding the maximum likelihood estimate of their (conditional) probabilities. In practice, for large $n$, even large training data cannot contain all possible sequences, so the $n$-gram model will contain a considerable amount of zero transition probabilities. Since this would lead to problems in, e.g., a speech recognition system, those entries are increased by a small constant to \emph{smooth} the model, for example using Laplace's law~\cite[pp.~202]{Manning_NLP}.

\begin{figure}
	\centering
	\includegraphics[width=0.35\textwidth]{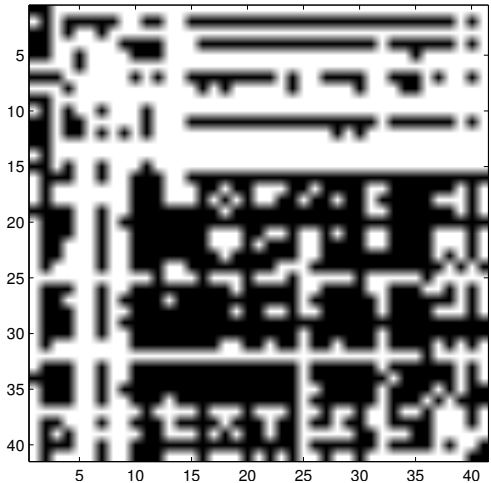}
	\caption{The adjacency matrix of the bi-gram model of ``The Great Gatsby''. The first two states are line break (LB) and space (\charl{ }), followed by punctuations. The block in the lower right corner indicates interactions of letters and punctuation following letters.}
	\label{fig:adjacencyMatrix}
\end{figure}

Since, by Proposition~\ref{prop:Msfs2}, an information-preserving lumping is more efficient for a sparse transition matrix, we refrain from smoothing and use the maximum likelihood estimates of the model parameters instead. We trained a bi-gram letter model of F. Scott Fitzgerald's \emph{``The Great Gatsby''}, a text containing roughly 270000 letters. To reduce the alphabet size and, thus, the run-time of the algorithm, we replaced all numbers by \charl{\#} and all upper case by lower case letters. We left punctuations unchanged, yielding a total alphabet size of $N=41$. The adjacency matrix of the bi-gram model can be seen in Fig.~\ref{fig:adjacencyMatrix}; the maximum out-degree of the Markov chain is 37.

Of the 820 possible merges only 21 are admissible. Furthermore, there are 129, 246, and 90 $\mathsf{SFS}(2)$-lumpings to sets of cardinalities 39, 38, and 37, respectively. Only two triples can be merged, namely $\{$LB, \charl{\$}, \charl{x}$\}$ and $\{$LB, \charl{(}, \charl{x}$\}$, where LB denotes the line break. Of the more notable pair-wise merges we mention $\{$\charl{(},\charl{)}$\}$, $\{$\charl{(},\charl{z}$\}$, and the merges of \charl{\#} with colon, semicolon, and exclamation mark. Especially the first is intuitive, since parentheses can be exchanged to, e.g., \charl{|} while preserving the meaning of the symbol\ifthenelse{\arxiv=1}{\footnote{Whether the symbol initiates or terminates a parenthetic expression is determined by whether the symbol is preceded or succeeded by a blank space. Unless parenthetic expressions are nested,  simple counting distinguishes between initiation and termination.}.}{.}

Finally, we determined the lumping yielding maximum compression, i.e., the one for which $\ent{Y}$ is minimal. This lumping, merging $\{$LB, \charl{\$}, \charl{x}$\}$, {$\{$\charl{!}, \charl{\#}$\}$}, and {$\{$\charl{(}, \charl{,}$\}$}, decreases the entropy from 4.3100 to 4.3044 bits. These entropies roughly correspond to the 4.03 bits derived for Shannon's first-order model, which contains only 27 symbols~\cite[p.~170]{Cover_Information2}.


Without preprocessing, an exhaustive search is significantly more expensive: With an alphabet size of $N=77$ and 357 admissible pairs, the first iteration of the algorithm already checks roughly 60000 \emph{distinct} partitions with 75 elements, most of them satisfying the $\mathsf{SFS}(2)$-condition. Proposition~\ref{prop:Msfs2} yields $M \ge 65$.

A modified algorithm only retains the best (in terms of the entropy of the marginal distribution) partition in each iteration. This greedy heuristic achieves a compression from 4.5706 to 4.4596 bits with $M=66$. We do not know if in this example $M=65$ can be attained, even by an exhaustive search.

Finally, we trained a tri-gram model of the same text, without preprocessing the alphabet, and lifted the resulting second-order Markov chain to a first-order Markov chain on $\dom{X}^2$ (states are now letter tuples). Eliminating all non-occurring tuples reduces the alphabet size to $N=1347$. Proposition~\ref{prop:Msfs2} yields $M\ge 53$. There are roughly 900000 admissible pairs. A further modified algorithm, considering only 10 random admissible pairs in each iteration, achieves a compression from 8.0285 to 7.1781 bits. The algorithm terminates at $M=579$. The question, whether a reduction to $M<77$ is possible (thus replacing a second-order Markov model by one on a smaller state space) remains open.

\section{Conclusion}
We presented a sufficient condition for merging states of a Markov chain such that the resulting process is second-order Markov and has full model information. We furthermore developed an iterative algorithm finding all such merges for a given transition matrix. Finally, we presented a lower bound on the cardinality of the reduced state space depending on the maximum out-degree of the associated transition graph.

The application of our algorithm to a bi-gram letter model suggests its practical relevance for model-order reduction. Future work shall investigate whether it can be successfully applied to $n$-gram models ($n>2$) and whether it is asymptotically optimal.

\section*{Acknowledgments}
The authors gratefully acknowledge Franz Pernkopf, Signal Processing and Speech Communication Laboratory, Graz University of Technology, for suggesting $n$-grams as a possible application of their theoretic results.

\appendix

\section{Proofs}
\subsection{Proof of Proposition~\ref{prop:TrajPres}}
We recall from~\cite{GeigerTemmel_kLump} that $\entrate{X|Y}=0$ implies, for all $n$,
\begin{multline}
	\forall\check{x},\hat{x}\in\dom{X}, \yvec\in\dom{Y}^{n-2}{:}\\
	\Prob{X_1=\check{x},Y_2^{n-1}=\yvec,X_{n}=\hat{x}}>0\\
	\Rightarrow \exists! \xvec\in\dom{X}^{n-2}{:}\\\ \Prob{X_2^{n-1}=\xvec|X_1=\check{x},Y_2^{n-1}=\yvec,X_{n}=\hat{x}}=1\,.\label{eq:preimageProperty}
\end{multline}
We thus obtain a bound on a realization $t_n$ of the preimage count (i.e., for $Y_1^n=\yvec$)
 \begin{align*}
	&t_n \\
	&=\sum_{\xvec\in\preim{\yvec}}\left[\Prob{X_1^n=\xvec}>0\right]\\
	&=\sum_{\xvec\in\dom{X}^n}\left[\Prob{X_1^n=\xvec|Y_1^n=\yvec}>0\right]\\
	&=\sum_{\xvec\in\dom{X}^n}\left[\Prob{X_1=x_1,X_n=x_n|Y_1^{n}=\yvec}>0\right]\\
	&\quad \times \left[\Prob{X_2^{n-1}=x_2^{n-1}|Y_1^n=\yvec,X_1=x_1,X_n=x_n}>0\right]\\
	&\stackrel{(a)}{=}\sum_{\substack{x_1\in\preim{y_1}\\ x_n\in\preim{y_n}}}\left[\Prob{X_1=x_1,X_n=x_n|Y_1^{n}=\yvec}>0\right]\\
	&\le N^2<\infty
\end{align*}

where $(a)$ is due to~\eqref{eq:preimageProperty}. Since this holds for all $n$ and all realizations, this proves
\begin{equation}
	\entrate{X|Y}=0 \Rightarrow \exists\, C<\infty{:}\ \sup_{n\to\infty}\preimCnt\le C
\end{equation}
With~\eqref{eq:OrigTrajPreservation}, the reverse implication is trivial.
\endproof
\subsection{Proof of Corollary~\ref{cor:Mbound}}
The proof employs elementary results from graph theory: Let $\Amat$ denote the adjacency matrix of the Markov chain, i.e., $A_{i,j}=[P_{i,j}>0]$. The number of closed walks of length $k$ on the graph determined by $\Amat$ is given as~\cite[p.~24]{Cvetkovic_Graphs}
\begin{equation}
	\sum_{i=1}^N \lambda_i^k
\end{equation}
where $\{\lambda_i\}_{i=1}^N$ is the set of eigenvalues of $\Amat$.

Let $t_X^k$ denote the number of sequences $\xvec\in\dom{X}^k$ of $\Xvec$ with positive probability, i.e.,
\begin{equation}
	t_X^k = \sum_{\xvec\in\dom{X}^k} \left[\Prob{X_1^k=\xvec}>0\right]\,.
\end{equation}
Clearly, $t_X^k\ge \sum_{i=1}^N \lambda_i^k$. Furthermore, defining $t_Y^k$ similarily we obtain $t_Y^k\le M^k$. With $\lambda_{\max}$ denoting the largest eigenvalue of $\Amat$,
\begin{equation}
	\frac{t_X^k}{t_Y^k} \ge \frac{\sum_{i=1}^N \lambda_i^k}{M^k}\ge \left(\frac{\lambda_{\max}}{M}\right)^k\,.
\end{equation}
If $\lambda_{\max}>M$, then the ratio of possible length-$k$ sequences of $\Xvec$ to those of $\Yvec$ increases exponentially. Then, the \emph{pigeon-hole-principle} implies that also the preimage count $\preimCnt$ is unbounded. Thus,
\begin{equation}
	\entrate{X|Y}=0 \Rightarrow M\ge\lambda_{\max}\,.
\end{equation}
Finally, the \emph{Perron-Frobenius theorem} for non-negative matrices~\cite[Cor.~8.3.3]{Horn_Matrix} bounds the largest eigenvalue of $\Amat$ from below by the minimum out-degree of $\Pvec$.
\endproof
\subsection{Proof of Proposition~\ref{prop:SFSRefined}}
We prove the proposition by contradiction: Assume $(\Pvec,h)$ violates $\mathsf{SFS}(k)$. Then there exists a $\zvec\in\dom{Z}^{k-1}, z\in\dom{Z}$ such that there exist two distinct $\xvec',\xvec''\in h^{-1}[\zvec]$ and two, not necessarily distinct $x',x''\in h^{-1}[z]$ such that
\begin{align}
	\Prob{X_2^k=\xvec'|Z_2^n=\zvec,X_1=x'}&>0
\intertext{and}
	\Prob{X_2^k=\xvec''|Z_2^n=\zvec,X_1=x''}&>0\,.
\end{align}
In other words, there are two different sequences $\xvec',\xvec''$ accessible from either the same ($x'=x''$) or from different ($x'\neq x''$) starting states. 

Now take $\yvec=f(\zvec)$ and $y=f(z)$. Since $h$ is a refinement of $g$, we have $h^{-1}[\zvec]\subseteq\preim{\yvec}$ and $h^{-1}[z]\subseteq\preim{y}$. As a consequence, $\xvec',\xvec''\in\preim{\yvec}$ and $x',x''\in \preim{y}$, implying that $(\Pvec,g)$ violates $\mathsf{SFS}(k)$. This proves
\begin{equation}
	(\Pvec,h) \text{ violates } \mathsf{SFS}(k) \Rightarrow (\Pvec,g) \text{ violates } \mathsf{SFS}(k)\,.
\end{equation}
The negation of these statements completes the proof.
\endproof

\bibliographystyle{IEEEtran}
\bibliography{IEEEabrv,references}

\end{document}